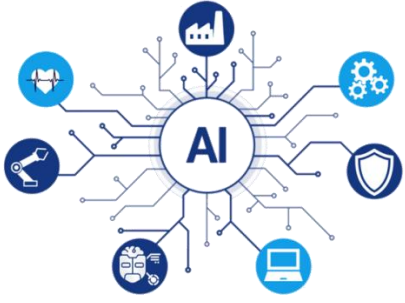

# Turkish Voice Commands based Chess Game using Gammatone Cepstral Coefficients

Gizem KARACA[1*], Yakup KUTLU[1],

[1]Department of Computer Engineering, Iskenderun Technical University, Hatay, TURKEY

**Abstract**

This study was carried out to enable individuals with limited mobility skills to play chess in real time and to play games with the individuals around them without being under any social distress or stress. Voice recordings were taken from 50 people (23 men and 27 women). While recording the sound, 29 words from each person were used which are determined as necessary for playing the game. Mel Frequency Coefficients (MFCC) and Gammatone Cepstral Coefficients (GTCC) qualification methods were used. In addition, k-NN, Naive Bayes and Neural Network classification methods were used for classification. Two different classification procedures were applied, namely, person-based and general. While the performance rate in person-based classification ranged from 75% to 98%, a performance over 84% was achieved in general classification.

## 1. INTRODUCTION

People cannot go out when it is not compulsory and even if they go out the fact that they have to meet with their family or friends with a certain social distance causes people to feel psychologically and socially incomplete. In this study, chess, which is also used in the treatment of some diseases such as alzhemir, focusing problem, forgetfulness, and substance addiction, has been tried to be done in today's world where people can socialize with their family and friends and relieve stress by playing games.

Although chess has remained in the foreground and sometimes in the background since ancient times, it has not lost its value and is used as the subject of many studies in the literature (Janko, 2016; Newell, 1988; Nabiyev, 2015; Hajari, 2014). It is a brain game that will be used by many studies even years later and also continue to be played as a game.

It was aimed to create a platform where the person can play chess with Turkish voice commands. Voice recordings are classified by different processes and according to that chess is played on the interface in real time with the results obtained. It is very important for the classification of sound signals in order to play the game quickly and accurately with voice commands. Therefore, the methods that give the fastest and best results with different methods have been tried to be determined. Voice recordings were taken from people beforehand, or a platform was created where voice recording can be made before the person plays a game.

### 1.1. Precious Studies

*Corresponding Author: Gizem Karaca, E-mail: krc.gizem343@gmail.com*

Automatic speech recognizers perform poorly in noisy environments. However, people can perceive speech and understand what is wanted to be said, even in noisy environments. Shao et al. (2009) created a more advanced speech recognition based on people's perception of speech. It was stated that a better result was obtained in this study, which was created using the GTCC method.

Valero & Alias(2012) used MFCC and GTCC methods in their study for the recognition and classification of non-speech sound. It has been tried to determine which method performs better in the recognition and classification of sound by comparing. As a result of the study, it was determined that the results obtained with the GTCC method were more successful.

Voice fraud is one of the most important challenges in speech verification (SV) systems. Das et al. (2016), developed a system that can distinguish voice fraud in their study. For this, MFCC, GTCC, modified gammatone frequency cepstral coefficients (MGFCC) and cosine normalized phase cepstral coefficients (CNPCC) methods were used.

The analysis of the sound signal, the correct recognition of the sound and its digital processing at this time constitute the main structure of the systems that are created for the automatic recognition of the sound. In the study, MFCC and GTCC methods were compared (Fathima & Raseena, 2013).

A study has been made for in which individuals with limited mobility can play chess with voice commands. Voice recordings from 30 people were feature selection using the MFCC method, and classified with k-NN method (Kutlu & Karaca,2019).

This study consists of three main modules. The first of these is the user module, in this module, voice recordings which is consist of 23 male and 27 female and in total 50 people were firstly recorded. A total of 14,000 sound recordings were taken with 290 voice recordings by each person. Sound recordings are first preprocessed by clearing them from noise and unnecessary words. Then, MFCC and GTCC methods are used for the feature selection of the signals. The selected attributes are classified using k-NN methods. Two different procedures are applied to determine the person who encompasses all the data in this classification process. As a result of the classification of all data, a performance close to 80-85% is obtained in general. In determining the person, a success between 75% and 98% has been achieved. It was determined that the method of determining the voice recording of the person gave a better result. Therefore, it is planned to develop three different methods for playing games. The first method is to determine the person whose voice has been recorded by looking at the database and to ensure that the person plays a game from those voices. Second method is that the person's voice recording is not taken in advance and the voice recording is added to the database in real time, and then the person can play games based on voice recordings. The final method is that to ensure they can play the game in case the person do not have the voice recording or if the person do not want their voice to be recorded. The second module is called the chess module and in this module it consists of the chess interface. The interface is provided in this module while playing chess game. The game has functions that enable the person to play chess against the computer or two people mutually. The third module was created for the interaction of the previous two modules. It is called the human computer interaction module. After the received voice recordings are classified in the user module, the necessary connections have been created to ensure the connection with the chess module and to make the desired move of the received sound signal on the interface. With the operation of this module, chess is played in real time.

**2. MATERIALS AND METHODS**

The voice commands that are 50 people (23 male and 27 female) "a, b, c, d, e, f, g, h, 1, 2, 3, 4, 5, 6, 7, 8, kale, at, fil, vezir, şah, piyon, mat, rook, başla, yeni oyun, çekil, geri al and kapat" have been recorded. While recording the voice, it is checked whether the person has any illness such as flu, sore throat or cough. Because the voice recording of a person similar to these diseases may be different due to a broken or changed voice, and if the same person wants to play with this voice recording, his voice may not be recognized correctly. Therefore, it is checked by asking the person first. While each person's voice is recorded, 10 recordings of each word are taken each consist of one second. In total 290 sound recordings were taken from each person and 14.000 sound recordings in total. The recorded voices were used after pre-processing. The sound recordings that were imported into the database were first converted into a wav sound file so that they could be used. It is then pre-processed to remove the noise and unnecessary sounds that may be present in the sound recording. After the pretreatment, the best method was tried to be determined by using MFCC and GTCC methods for the feature selection of sound recordings. For this, classification methods come into play. The results obtained in the methods such as k-NN, the results of two different methods were evaluated and the method that gave the best result was used.

In addition, a function that allows volunteers to add voice recordings to the database in real time has been created. Voice recordings of the words which are determined were taken from the people who did not have a voice recording before with the person-based calibration function. Here, the voice recording of 29 words which are determined by the given commands are taken for the person who do not have a recording in the database. In case the given recording contains different words or is taken as blank, it was tried to ensure that the correct sound recording was obtained by warning the person. Since the sound recordings taken here are taken directly from the wav sound file type, the sound recordings are first pre-processed and then classified using the determined methods. Then the person can play a game.

**2.1. Pre- Processing**
The sound recordings saved in the database before pre-processing are converted into the appropriate format. In this study, it was determined that the recorded sounds were converted to wav sound file and the sound recordings were converted into wav sound file. Later, while recording the voice, unnecessary words and sounds that the person makes voluntarily or unintentionally, noise and sounds originating from the environment are cut and processed. This process is applied to all voice recordings given by the person.

**2.2. Attribute Extraction and Classification**
The pre-processed sound recordings are first applied to the feature selection process. And the results obtained with the feature selection are used for the classification process. Therefore, in order to determine the method that gives the best results and to make the classification correctly, different methods have been used to obtain results. While the classification process is carried out, the results obtained by feature selection are classified using the determined methods. In this study, MFCC and GTCC methods were applied for the selection of attributes. In the classification process, k-NN, Naive Bayes and SVM methods were used and the method that gave the best and most accurate results was determined.

It is used for multiple purposes on audio files with MFCC. As an example, it is used as a feature extraction method in sound recordings for certain purposes such as distinguishing the sounds determined from audio files, determining gender and speaker. (Chakraborty & Parekh, 2018). In this study, it is used to determine the speaker's voice and to classify the determined voices.

In the MFCC method, the framing process is applied first to the pre-processed audio signals, and after this process, the uncertainty between the frames is tried to be removed by the windowing process. After these operations, fourier transform of audio signals is applied. And mel is passed through the frequency filter. After the filtering process, MFCC values consisting of 14 coefficents are obtained by applying inverse fourier transformation.

The GTCC method calculates the coefficient similar to the MFCC method. The difference between the methods is the filtering method they use. While the mel frequency filter is used in MFCC, a gammatone filter is applied in GTCC. Other than that, the transactions and process steps are the same (Fathima, 2013; Das, 2016; Valero, 2012; Shao, 2009).

**2.2.1. K Nearest Neighbor Classification**
Nearest neighbor (k-NN) is a simple classification method. However, it is a frequently used classification method due to the good results obtained while classifying with k-NN and the rapid results. (Ahmad, 2016; Chakraborty, 2018).

It is one of the methods used in this study. Classification tables containing more than one class of data are obtained by controlling K-NN distance intervals, and the accuracy values of these tables are evaluated using the sensitivity and selectivity functions (Yayik, 2015; Altan, 2016; Kutlu, 2019):

$$SEN = \frac{TP}{TP+FN} x 100\%$$

$$SEL = \frac{TP}{TP+FP} x 100\%$$

$$Genel\ Basarim = \frac{TP+TN}{TP+FN+TP+FP} x 100\%$$

**3. RESULTS and DISCUSSION**
It has been determined that the results obtained with different methods have a better performance of the GTCC method. When the general classification and person-based classification were evaluated, it was determined that person-based results gave better results for MFCC and GTCC. Overall GTCC achieved a performance of over 84%, but an average of over 90% in person-based classification.

**Tablo1. Avarage Result of Person-Based Classification GTCC Method with k-NN**

| Subject | SEN | SEL | SPE |
|---|---|---|---|
| *1* | 88.87 | 88.62 | 99.60 |
| *2* | 88.12 | 87.59 | 99.56 |
| *3* | 85.81 | 85.17 | 99.47 |
| *4* | 96.49 | 96.21 | 99.87 |
| *5* | 88.96 | 87.93 | 99.57 |
| *6* | 92.06 | 91.38 | 99.69 |
| *7* | 81.72 | 81.38 | 99.34 |
| *8* | 95.32 | 94.83 | 99.82 |
| *9* | 88.05 | 86.90 | 99.53 |
| *10* | 86.76 | 86.21 | 99.51 |
| *11* | 93.42 | 92.76 | 99.74 |
| *12* | 78.90 | 78.62 | 99.24 |
| *13* | 94.39 | 94.14 | 99.79 |
| *14* | 94.08 | 93.79 | 99.78 |
| *15* | 93.08 | 92.07 | 99.72 |
| *16* | 88.96 | 88.28 | 99.58 |
| *17* | 98.51 | 98.28 | 99.94 |
| *18* | 90.86 | 90.34 | 99.66 |
| *19* | 93.85 | 93.10 | 99.76 |
| *20* | 98.36 | 98.28 | 99.94 |
| *21* | 94.55 | 93.45 | 99.77 |
| *22* | 93.19 | 92.41 | 99.73 |
| *23* | 92.77 | 92.07 | 99.72 |
| *24* | 95.46 | 95.17 | 99.83 |
| *25* | 87.48 | 86.21 | 99.51 |
| *26* | 85.63 | 85.17 | 99.47 |
| *27* | 93.03 | 92.41 | 99.73 |
| *28* | 86.60 | 84.83 | 99.46 |
| *29* | 79.71 | 78.62 | 99.24 |
| *30* | 88.74 | 87.93 | 99.57 |
| *Average* | 90.18 | 89.80 | 99.64 |